\documentclass[aps,eqsecnum,preprint,preprintnumbers,12pt]{revtex4}
\usepackage{epsfig,psfrag}

\def\nn{\nonumber}

\begin{document}

\title{Instanton Determinant with Arbitrary Quark Mass: WKB Phase-shift Method and Derivative Expansion}
\author{Gerald V. Dunne}\email{dunne@phys.uconn.edu}
\affiliation{Department of Physics, University of Connecticut, Storrs, CT 06269, USA\\
CSSM, University of Adelaide, SA 5005, Australia}
\author{Jin Hur} \email{hurjin2@snu.ac.kr}
\author{Choonkyu Lee} \email{cklee@phya.snu.ac.kr}
\affiliation{Department of Physics and Center for Theoretical
Physics\\ Seoul National University, Seoul 151-742, Korea}
\author{Hyunsoo Min} \email{hsmin@dirac.uos.ac.kr}
\affiliation{Department of Physics, University of Seoul, Seoul 130-743, Korea\\ Department of Physics, University of Connecticut, Storrs, CT 06269, USA}

\begin{abstract}
The fermion determinant in an instanton background for a quark
field of arbitrary mass is studied using the Schwinger proper-time
representation with WKB scattering phase shifts for the relevant
partial-wave differential operators. Previously, results have been
obtained only for the extreme small and large quark mass limits,
not for intermediate interpolating mass values. We show that
consistent renormalization and large-mass asymptotics 
requires up
to third-order in the WKB approximation. This procedure leads to
an almost analytic answer, requiring only modest numerical
approximation, and yields excellent agreement with the well-known
extreme small and large mass limits. We estimate that it differs
from the exact answer by no more than 6\% for generic mass values.
In the philosophy of the derivative expansion the same amplitude
is then studied using a Heisenberg-Euler-type effective action,
and the leading order approximation gives a surprisingly accurate
answer for all masses.
\end{abstract}
\maketitle

\section{Introduction}

The one-loop effective action plays a central role in quantum field theory.
For a specific background field it represents quantum effects of direct physical relevance
(e.g., the induced vacuum energy in the given background), while for general background fields it corresponds to the proper-vertex generating functional of the theory. The ultraviolet
divergence or renormalization problem for the one-loop effective action is
well-understood, but it is notoriously difficult to evaluate explicitly its full
finite part in the presence of a given
nontrivial background. For nonabelian theories in three spatial dimensions, there are
essentially only two types of exact calculations for this
\emph{full} amplitude -- the Heisenberg-Euler-type nonlinear
action describing vacuum polarization and pair creation effects in
the background of (covariantly) constant gauge field strengths \cite{heisenberg,
schwinger, brownduff,dunne}, 
and the QCD one-loop effective action for massless quarks in a
self-dual Yang-Mills background field (describing
instantons) \cite{thooft, brown}.

To study instanton-related physics it is of fundamental importance
to determine the one-loop tunnelling amplitude given by the
Euclidean one-loop effective action in the background of a single
(anti-)instanton \cite{belavin}. 't Hooft \cite{thooft} succeeded in
calculating the corresponding contribution by massless scalar or
quark fields exactly in the 1970s, but such an exact calculation
is no longer possible if the fields have non-zero mass.
For various phenomenological applications \cite{schaffer}
(and also for the extrapolation of lattice results \cite{lattice},
obtained at unphysically large quark masses, to lower physical
masses), it is important to have more definite information about
the contributions due to quarks of not-so-small mass. In this
work, we will describe our approach to this problem and present
explicit numerical results.

Previously, two of us (with Kwon) \cite{kwon} studied the
corresponding one-loop effective action with arbitrary mass, using
a smooth interpolation between the results obtained in the large
and small mass limits. The expression in the large mass
limit is naturally obtained from the Schwinger-DeWitt expansion \cite{dewitt, lee,shifman} (or the
heat-kernel expansion) within the proper-time representation of
the effective action, while the result in the
small mass limit follows from the works of 't Hooft \cite{thooft},
Carlitz and Creamer \cite{carlitz}, and Kwon et al. \cite{kwon}. In
this work, we use a systematic approximation
for any quark mass -- the effective action for any mass is
calculated approximately without invoking an ad hoc interpolation
procedure. The basic strategy is as follows. We use the
proper-time representation of the effective action, which requires the
explicit functional form of
\begin{equation}
F(s)=\int d^4 x \;\textrm{tr}\langle x|{e^{-s(-{\rm D}^2
)}-e^{-s(-\partial^2 )}}|x\rangle
\label{propertime}
\end{equation}
The instanton background enters through the covariant derivative
$D_{\mu}$ in $D^2 \equiv D_{\mu}D_{\mu}$, and this function $F(s)$ can
be expressed in terms of scattering phase shifts \cite{thooft,wasson,
moss}
for the partial-wave quadratic differential operators
related to $D^2$. We evaluate these phase shifts in the
quantum-mechanical WKB approximation,
extended to the third-order correction terms \cite{dunham, bender}
to ensure the correct small-$s$ behavior for $F(s)$. This leads to a numerical
expression for $F(s)$, which in turn yields the effective action for
arbitrary mass value. The resulting mass dependence is fully
consistent with the conjecture made by Kwon et al \cite{kwon}, on
the basis of the explicitly known results at the opposite ends.

Our partial-wave WKB phase-shift method should provide a practical
approximation scheme for the one-loop effective action in a broad
class of background fields. In this approach, $F(s)$ generally
takes a \emph{local} form in the potentials and their derivatives
-- see (\ref{local}) below. In the light of this observation,
we
conclude this paper by comparing to another method, the derivative
expansion, which produces such local expressions. Surprisingly,
the leading order of the derivative expansion provides a
remarkably good approximation for general values of the quark
mass.

\section{Effective action, proper-time representation, and phase
shifts} \label{sec2}

Due to a hidden supersymmetry of the system with
 a quark in a
background instanton field, the one-loop effective action of a
Dirac spinor field of mass $m$ (and isospin $\frac{1}{2}$),
$\Gamma^{F}(A;m)$, can be related to the corresponding scalar
effective action (for a complex scalar of mass $m$  and isospin
$\frac{1}{2}$) by \cite{thooft,brown, kwon}
\begin{equation}
\Gamma^{F}(A;m)= -\frac{1}{2}\ln\left(\frac{m^2}{\mu^2}\right)
-2\, \Gamma^{S}(A;m),
\label{susy}
\end{equation}
The first contribution corresponds to the existence of a zero
eigenvalue in the spectrum of the Dirac operator for a single
instanton background. This relationship (which is special to a
self-dual background) 
 has the important consequence that it is
sufficient to consider the scalar effective action
$\Gamma^{S}(A;m)$ to learn also about the corresponding fermion
effective $\Gamma^{F}(A;m)$, for any mass value m.

We consider an SU(2) single instanton background \cite{thooft}
\begin{eqnarray}
&& A_{\mu}(x) \equiv A_{\mu}^{a}(x)\frac{\tau^{a}}{2}= \frac{\eta_{\mu\nu
a}\tau^{a}x_{\nu}}{r^2+\rho^2},\;\;(\mu=1,2,3,4;\;r \equiv
\sqrt{x_{\mu}x_{\mu}}) \label{insta} \\
&& F_{\mu\nu}(x) \equiv F_{\mu\nu}^{a}(x)\frac{\tau^{a}}{2}
=-\frac{2\rho^2 \eta_{\mu\nu a}\tau^{a}}{(r^2+\rho^2)^2},
\label{instf}
\end{eqnarray}
The regularized one-loop scalar effective action has  the proper-time representation
\begin{eqnarray}
\Gamma_{\Lambda}^{S}(A;m) &=& - \int_{0}^{\infty}
\frac{ds}{s}(e^{-m^2 s}-e^{-\Lambda^2 s}) \int d^4
x\;\textrm{tr}\langle x|{e^{-s(-{\rm D}^2
)}-e^{-s(-\partial^2 )}}|x\rangle \nonumber \\
&\equiv& - \int_{0}^{\infty} \frac{ds}{s}(e^{-m^2 s}-e^{-\Lambda^2
s}) F(s),
 \label{ptaction}
\end{eqnarray}
where $D^2 \equiv D_{\mu}D_{\mu}$ with
$D_{\mu}=\partial_{\mu}-iA_{\mu}(x)$. From this one obtains the
renormalized effective action, in the minimal subtraction scheme,
as
\begin{eqnarray}
\Gamma^{S}(A;m) &=& \lim_{\Lambda\rightarrow\infty}
\left[\Gamma_{\Lambda}^{S}(A;m)-\frac{1}{12} \frac{1}{(4\pi)^2}
\ln \left(\frac{\Lambda^2}{\mu^2}\right) \int d^4
x\;\textrm{tr}(F_{\mu\nu}F_{\mu\nu})\right] \nonumber \\
&\equiv& \lim_{\Lambda\rightarrow\infty}
\left[\Gamma_{\Lambda}^{S}(A;m)-\frac{1}{12}
\ln\left(\frac{\Lambda^2}{\mu^2}\right)\right].
\label{renaction}
\end{eqnarray}
Moreover, by dimensional considerations, we may
introduce the modified scalar effective action
$\tilde{\Gamma}^{S}(m\rho)$, which is a function of $m\rho$ only, defined by
\begin{equation}
\Gamma^{S}(A;m)= \frac{1}{6}\ln(\mu\rho)+\tilde{\Gamma}^{S}(m\rho)
\label{modaction}
\end{equation}
and concentrate on studying the $m\rho$ dependence of
$\tilde{\Gamma}^{S}(m\rho)$. Then there is
no loss of generality in our setting the instanton scale $\rho=1$ henceforth.

The small-$s$ behavior of $F(s)$, as given by the Schwinger-DeWitt
expansion, reads \cite{kwon}
\begin{equation}
s\rightarrow 0+ \quad :\quad F(s)\sim -\frac{1}{12}+\frac{1}{75}s+\frac{17}{735}s^2-\frac{116}{2835}s^3+\cdots
\label{smalls}
\end{equation}
Insering this result into (\ref{ptaction}) gives rise to the
following large-mass expansion of  $\tilde{\Gamma}^{S}(m)$:
\begin{equation}
m\rightarrow \infty \;:\;\tilde{\Gamma}^{S}(m)= -\frac{1}{6}\ln
m-\frac{1}{75 m^2}-\frac{17}{735 m^4}+\frac{232}{2835 m^6}+\cdots.
\label{largemass}
\end{equation}
Note that the Schwinger-DeWitt expansion is a small $s$ expansion
for $F(s)$, which naturally leads to a large $m$ expansion for
$\tilde{\Gamma}^{S}(m)$. To obtain an expression for
$\tilde{\Gamma}^{S}(m)$ for general values of $m$ we need a more
general expression for $F(s)$. In the small mass limit, on the
other hand, a completely independent calculation for
$\tilde{\Gamma}^{S}(m)$ has been given \cite{thooft, kwon,
carlitz}, based on the fact that the massless propagators in an
instanton background are known in closed-form:
\begin{equation}
m\rightarrow 0 \quad :\quad \tilde{\Gamma}^{S}(m)=\alpha\left(\frac{1}{2}\right)+\frac{1}{2}(\ln m+\gamma-\ln 2)m^2
+O(m^4),
 \label{smallmass}
\end{equation}
where $\alpha(\frac{1}{2})\simeq 0.145873$, and $\gamma\simeq
0.5772\dots$ 
is Euler's constant. From this small $m$ behavior it
is possible to infer that $F(s)\sim -\frac{1}{4s}$ as
$s\rightarrow\infty$. However, this information does not determine
the magnitude of $\tilde{\Gamma}^{S}(m)$ even for small $m$, since
the integral in (\ref{ptaction}) is dominated by contributions
from non-asymptotic $s$-values.

In this paper we make use of the fact that the function $F(s)$ may be expressed in terms of appropriate
scattering phase shifts. Note that the differential
operator $-D^2$ in the instanton background (\ref{insta}) (with
$\rho=1$) can be cast in the form \cite{thooft}
\begin{equation}
-D^2 =-\frac{\partial^2}{\partial
r^2}-\frac{3}{r}\frac{\partial}{\partial r}
+\frac{4}{r^2}\vec{L}^2 +\frac{4}{r^2+1}(\vec{J}^2 -\vec{L}^2
)-\frac{4}{(r^2 +1)^2}\vec{T}^2,
 \label{instop}
\end{equation}
where $\vec{T}^2 \equiv T^a T^a$, and eigenvalue $T^2 =t(t+1)\displaystyle \frac{3}{4}$ appropriate to isospin $t=\displaystyle
\frac{1}{2}$; $\vec{L}^2 \equiv L_a L_a$ with $L_a -\frac{i}{2}\eta_{\mu\nu a}x_{\mu}\partial_{\nu}$ (satisfying
angular-momentum commutation relations) and eigenvalues $L^2 l(l+1)$, $l=0,\;\displaystyle
\frac{1}{2},\;1,\;\frac{3}{2},\;\cdots\;$; $\vec{J}^2 \equiv
(\vec{L}+\vec{T})^2$ with eigenvalues $J^2 =j(j+1)$
, $j=|\;l\pm
t\;|=|\;l \pm \displaystyle \frac{1}{2}\;|$. Without any
background, we have the differential operator
\begin{equation}
-\partial^2 = - \frac{\partial^2}{\partial r^2} -
\frac{3}{r}\frac{\partial}{\partial r}+\frac{4}{r^2}\vec{L}^2,
\label{freeop}
\end{equation}
which corresponds to the t=0 case of the expression (\ref{instop}).
We may then consider the quantum mechanical scattering problem
with the Hamiltonian ${\cal H}= - D^{2} $, viz.,
\begin{equation}
{\cal H} \Psi \equiv \left[ - \frac{\partial^2}{\partial
r^2}-\frac{3}{r}\frac{\partial}{\partial
r}+\frac{4l(l+1)}{r^2}+\frac{4(j-l)(j+l+1)}{r^2+1}-\frac{3}{(r^2+1)^2}
\right] \Psi = k^2 \Psi
\label{insth}
\end{equation}
with the corresponding free Schr\"{o}dinger equation given by
\begin{equation}
{\cal H}_0 \Psi_0 \equiv \left[ - \frac{\partial^2}{\partial
r^2}-\frac{3}{r}\frac{\partial}{\partial r}+\frac{4l(l+1)}{r^2}
\right] \Psi_0 = k^2 \Psi_0 .
\label{freeh}
\end{equation}

As $r\rightarrow 0$, we assume that $\Psi$, $\Psi_0 \sim ({\rm
const.})\,r^{2l}$. Also, to make the spectrum discrete, it is
convenient to put the system in a large spherical box of radius
$R$, demanding a suitable boundary condition at $r=R$ (e.g., the
Dirichlet condition $\psi(r=R)=0$). Then the solutions of
(\ref{insth}) and (\ref{freeh}) have the asymptotic large-$r$
behaviors
\begin{eqnarray}
\Psi_{0n}(r) &\sim& 2C r^{-3/2}\cos[k_0 (n)(r+a)], \\
\Psi_{n}(r) &\sim& 2C r^{-3/2}\cos[k_0 (n)(r+a)+\eta(k(n))]
\label{larger}
\end{eqnarray}
where $\eta(k(n))$ denotes the related scattering phase shift, and the now discrete momenta
$k_0 (n)$, $k(n)$ ($n$: nonnegative
integers) satisfy the conditions \cite{thooft, moss}
\begin{eqnarray}
&& k(n+1)-k(n)=\frac{\pi}{R}+O\left(\frac{1}{R^2}\right)\;\;(=k_0
(n+1)-k_0 (n)), \nn\\
&& k_0 (n)=k(n) +
\frac{\eta(k(n))}{R}+O\left(\frac{1}{R^2}\right).
\label{momentum}
\end{eqnarray}
This scattering mode description may be considered for every
partial wave. If $[k^{l,j}(n)]^2$ and $[k_0^l (n)]^2$ denote the
energy eigenvalues introduced in association with (\ref{insth}) and
(\ref{freeh}), respectively, the function $F(s)$ (see (\ref{propertime}))
may then be represented as
\begin{equation}
F(s) = \sum_{l=0, \frac{1}{2},\cdots} \sum_j (2l+1)(2j+1)\sum_n
\left\{  e^{-s[k^{l,j}(n)]^2}- e^{-s[k_0^l (n)]^2} \right\},
\label{fexpansion}
\end{equation}
including the degeneracy factor $(2l+1)(2j+1)$. The
factor $(2j+1)$ corresponds to different eigenvalues of $J_3$,
while the factor $(2l+1)$ corresponds to the
eigenvalues of $\bar{L}_3$, the third component of the second set
of conserved angular-momentum $\bar{L}_a \equiv
-\frac{i}{2}\bar{\eta}_{\mu\nu a}x_{\mu}\partial_{\nu}$.

The phase-shift relations (\ref{momentum}) imply that for large $R$ we can write
\begin{equation}
e^{-s[k^{l,j}(n)]^2}- e^{-s[k_0^l
(n)]^2}=e^{-s[k^{l,j}(n)]^2}\left\{
\frac{2k^{l,j}(n)\eta_{l,j}(k(n))}{R}s+O\left(\frac{1}{R^2}\right)
\right\}.
\label{phasediff}
\end{equation}
Based on this observation, it is possible to replace the sum
$\sum_n$ in (\ref{fexpansion}) by an integral:
\begin{equation}
F(s)=\frac{2}{\pi}s\sum_{l=0,\frac{1}{2},\cdots}\sum_j
\int_0^{\infty} dk e^{-k^2 s}k(2l+1)(2j+1)\eta_{l,j}(k).
 \label{fexpression}
\end{equation}
Note that this expression
is not only an infinite series but also contains an improper
$k$-integral -- hence it must be considered 
carefully. In the instanton background in particular, the $l$-sum
and $j$-sum in (\ref{fexpression}) may not be considered in a
completely independent way. This follows from the nature of the
scattering problem as defined by (\ref{insth}) and (\ref{freeh});
according to the forms  for $\mathcal H$ and ${\mathcal H}_0$,
their small-$r$ behaviors match for a given $l$-value, but it is
the $j$-value that governs the large-$r$ behavior of the effective
potential in ${\mathcal H}$, and $j$ does not appear in ${\mathcal
H}_0$. This apparent mis-match can be resolved simply by
considering the phase shifts $\eta_{l,l+\frac{1}{2}}(k)$ and
$\eta_{l+\frac{1}{2},l}(k)$ (with the same associated degeneracy
factor) together as a package. With this understanding, the
expression (\ref{fexpression}) can now be cast in the form
\begin{equation}
F(s)=\frac{2}{\pi}s\sum_{l=0,\frac{1}{2},\cdots}\int_0^{\infty}dk
e^{-k^2
s}k(2l+1)(2l+2)\left\{\eta_{l,l+\frac{1}{2}}(k)+\eta_{l+\frac{1}{2},l}(k)
\right\}.
\label{ffinal}
\end{equation}
This form still requires careful treatment with regards to the $l$-sum and $k$-integration. We will argue below that the correct (gauge-invariant) procedure
is to have the various terms corresponding to the same 'energy'
eigenvalue (i.e., the same $k^2$-value) receive uniform
consideration. As a convenient check on this procedure, we confirm that the predicted small-$s$ behavior reproduces
the form in (\ref{smalls}).

\section{Systematic WKB phase-shift analysis}

To find the exact form of $F(s)$ with the help of (\ref{ffinal}),
one must first have complete knowledge of the scattering phase
shifts $\eta_{l,j}(k)$, and then carry out the needed infinite
sum/integration in a carefully controlled way. This is not possible in
general, and therefore one needs to develop a reliable approximation scheme to
determine the function $F(s)$. We provide such a scheme below,
which relies on a systematic WKB approximation for the
scattering phase shifts in question. In contrast to previous
applications \cite{wasson} of the WKB method for a similar purpose
(but in lower dimensions), the leading-order WKB result turns
out to be insufficient  even to ensure the correct
value for $F(s=0)$. [Recall that from (\ref{smalls}) we must have $F(s=0)=-\frac{1}{12}$ in order to
construct the renormalized effective
action as in (\ref{renaction})]. The necessity to include higher-order WKB contributions arises from the inaccuracy
of the $l$-sum in (\ref{ffinal}) with respect to large-$l$
contributions. While the degree of accuracy of the WKB
expression for the phase shifts is generally enhanced for higher
partial waves, the large degeneracy factor
$(2l+1)(2l+2)$ has the consequence that
leading-order WKB method is insufficient. Fortunately, as we shall see below, this can be remedied in a systematic way by including higher-order WKB correction terms.

Let us first consider the representation of $F(s)$ in the leading
WKB approximation. To that end, (\ref{insth}) may be rewritten as
an equation for $\bar{\Psi}(r)\equiv r^{3/2}\Psi(r)$:
\begin{equation}
\left\{ -\frac{\partial^2}{\partial
r^2}+\frac{4l(l+1)+\frac{3}{4}}{r^2}+\frac{4(j-l)(j+l+1)}{r^2
+1}-\frac{3}{(r^2 +1)^2} \right\}\bar{\Psi}(r) = k^2
\bar{\Psi}(r).
\label{insthmod}
\end{equation}
For this one-dimensional Schr\"{o}dinger-type equation, the
leading WKB solution can be given immediately, including the usual
Langer correction \cite{langer} to take into account the singular
centrifugal term. Phase shifts in the leading WKB approximation,
which can be extracted from this solution, read
\begin{equation}
\eta_{l,j}^{(1)}(k)= \int_{r_1 (k)}^{\infty}dr' \sqrt{k^2
-V^{l,j}(r')}- \int_{r_0 (k)}^{\infty}dr' \sqrt{k^2 -V_0^{l}(r')}
\label{leadingwkb}
\end{equation}
with
\begin{eqnarray}
V^{l,j}(r) &\equiv&
\frac{4(l+\frac{1}{2})^2}{r^2}+\frac{4(j-l)(j+l+1)}{r^2
+1}-\frac{3}{(r^2 +1)^2}, \label{pot} \\
V_0^{l}(r) &\equiv& \frac{4(l+\frac{1}{2})^2}{r^2}, \label{pot0}
\end{eqnarray}
In (\ref{leadingwkb}), $r_1 (k)$, $r_0 (k)$ denote the classical turning points
determined by the conditions $V^{l,j}(r_1 )=k^2$ and $V_0^l (r_0
)=k^2$, respectively.

We now define
\begin{equation}
X_{l,j}^{(1)}(s) \equiv \int_0^{\infty}dk\, e^{-k^2
s}\,k\,\eta_{l,j}^{(1)}(k)
\label{x1}
\end{equation}
as this kind of integral is relevant in the construction of $F(s)$ in
(\ref{ffinal}). Using the expression (\ref{leadingwkb}) for the phase
shift and changing the order of integrations, this quantity may
then be rewritten as
\begin{equation}
X_{l,j}^{(1)}(s) = \int_0^{\infty}dr \left[ \int_{k_1
(r)}^{\infty}dk\; e^{-k^2 s}k\sqrt{k^2 -V^{l,j}(r)}- \int_{k_0
(r)}^{\infty}dk\; e^{-k^2 s}k\sqrt{k^2 -V_0^{l}(r)} \right],
\label{x1wkb}
\end{equation}
where $k_1 (r)\equiv \sqrt{V^{l,j}(r)}$ and $k_0 (r)\equiv
\sqrt{V_0^{l}(r)}$. The $k$-integration in (\ref{x1wkb}) can be
carried out explicitly to give
\begin{equation}
X_{l,j}^{(1)}(s) = \frac{\sqrt{\pi}}{4s^{3/2}} \int_0^{\infty}dr
\left[ e^{-s V^{l,j}(r)}- e^{-s V_0^l (r)} \right].
\label{x1int}
\end{equation}
Thus, the leading
WKB expression for F(s) is:
\begin{eqnarray}
F^{(1)}(s)= \frac{1}{2\sqrt{\pi}\sqrt{s}} \int_0^{\infty}dr
\left(\sum_{l=0,\frac{1}{2},\cdots}(2l+1)(2l+2)\left\{ e^{-s
V^{l,l+\frac{1}{2}}(r)}-e^{-s V_0^{l}(r)} \right.\right. \nonumber \\
\left.\left. + e^{-s V^{l+\frac{1}{2},l}(r)}-e^{-s
V_0^{l+\frac{1}{2}}(r)} \right\} \right).
\label{fleading}
\end{eqnarray}

Higher-order WKB correction can also be included. For the
Schr\"{o}dinger equation (\ref{insthmod}), one can derive the
2nd-order and 3rd-order WKB phase shifts (i.e.,
$\eta_{l,j}^{(2)}(k)$ and $\eta_{l,j}^{(3)}(k)$), incorporating
the Langer correction in an appropriate manner, along the line
discussed in Refs. \cite{dunham, bender, langer}. Leaving the
somewhat involved details of this derivation
elsewhere \cite{preparation}, we shall here only report the results
for $X_{l,j}^{(2)}(s)$ and $X_{l,j}^{(3)}(s)$ (which are related
to the phase shift contributions of respective order by the
integral relation of (\ref{x1})):
\begin{eqnarray}
X_{l,j}^{(2)}&=& \frac{\sqrt{\pi}}{4s^{3/2}}\int_0^{\infty}dr
\left[ e^{-s V^{l,j}(r)} \left\{\frac{1}{4r^2 }s -\frac{1}{12}s^2
\frac{d^2 V^{l,j}(r)}{dr^2} \right\} \right. \nonumber \\
&& \hspace{6cm} \left. -e^{-s V_0^{l}(r)} \left\{\frac{1}{4r^2 }s
-\frac{1}{12}s^2
\frac{d^2 V_0^{l}(r)}{dr^2} \right\} \right],
\label{x2}\\
X_{l,j}^{(3)}&=& \frac{\sqrt{\pi}}{4s^{3/2}}\int_0^{\infty}dr
\left[ e^{-s V^{l,j}(r)} \left\{
\frac{5s^2}{32r^4}-\frac{s^3}{48r^2}\frac{d^2 V^{l,j}(r)}{dr^2} -
\frac{s^3}{288}\frac{d^4 V^{l,j}(r)}{dr^4}
+\frac{7s^4}{1440}\left( \frac{d^2 V^{l,j}(r)}{dr^2} \right)^2
\right\} \right. \nonumber \\
&&\;\;\;\;\; \left. -e^{-s V_0^{l}(r)} \left\{
\frac{5s^2}{32r^4}-\frac{s^3}{48r^2}\frac{d^2 V_0^{l}(r)}{dr^2} -
\frac{s^3}{288}\frac{d^4 V_0^{l}(r)}{dr^4}
+\frac{7s^4}{1440}\left( \frac{d^2 V_0^{l}(r)}{dr^2} \right)^2
\right\} \right].
\label{x3}
\end{eqnarray}
Inserting these results into  (\ref{ffinal}) we obtain the 3rd-order WKB expression for $F(s)$:
\begin{eqnarray}
F(s)_{\rm WKB} &=& F^{(1)}(s)+F^{(2)}(s)+F^{(3)}(s) \nonumber \\
&=& \frac{1}{2\sqrt{\pi}\sqrt{s}} \int_0^{\infty} dr \left[
\sum_{l=0,\frac{1}{2},\cdots}(2l+1)(2l+2)\left\{e^{-s
V^{l,l+\frac{1}{2}}(r)}H^{l,l+\frac{1}{2}}(r)-e^{-s
V_0^{l}(r)}H_0^{l}(r)
\right.\right. \nonumber \\
&&\hspace{4cm} \left.\left. +e^{-s
V^{l+\frac{1}{2},l}(r)}H^{l+\frac{1}{2},l}(r)-e^{-s
V_0^{l+\frac{1}{2}}(r)}H_0^{l+\frac{1}{2}}(r) \right\} \right]
\label{local}
\end{eqnarray}
where the $H(r)$ functions are local functions of the potentials
in (\ref{pot}) and (\ref{pot0}):
\begin{eqnarray}
H^{l,j}(r) &=& 1+\frac{s}{4r^2} +\frac{5s^2}{32r^4}
-\left(\frac{s^2}{12} +\frac{s^3}{48r^2}\right)\frac{d^2
V^{l,j}(r)}{dr^2} \nonumber \\
&& \hspace{4cm} -\frac{s^3}{288}\frac{d^4 V^{l,j}(r)}{dr^4}
+\frac{7s^4}{1440}\left(\frac{d^2 V^{l,j}(r)}{dr^2}\right)^2,
\label{hinst}\\
H_0^{l}(r) &=& 1+\frac{s}{4r^2} +\frac{5s^2}{32r^4}
-\left(\frac{s^2}{12} +\frac{s^3}{48r^2}\right)\frac{d^2
V_0^{l}(r)}{dr^2} \nonumber \\
&& \hspace{4cm}  -\frac{s^3}{288}\frac{d^4 V_0^{l}(r)}{dr^4}
+\frac{7s^4}{1440}\left(\frac{d^2 V_0^{l}(r)}{dr^2}\right)^2.
\label{hfree}
\end{eqnarray}

It is important that in evaluating (\ref{local}) the $l$-sum be done first, and then the $r$-integration.
This rule regarding what operations should be done
first follows  \cite{preparation} by adopting a definition of $F(s)$ as given by the infinite
'energy' cutoff limit (see the remark at the end of Sec.
\ref{sec2}). In fact, only with this procedure is the correct
small-$s$ behavior for $F(s)$ found (see below) -- this can be
taken as further (a posteriori) evidence for our rule. We have verified explicitly that carrying
out the $r$-integration for individual partial wave
contributions first and then summing leads to incorrect expressions. The difference for $F(s)$ is
$\frac{1}{4s}$, which leads to a spurious quadratic divergence
in the $s$-integral.

\begin{figure}[t]
\includegraphics{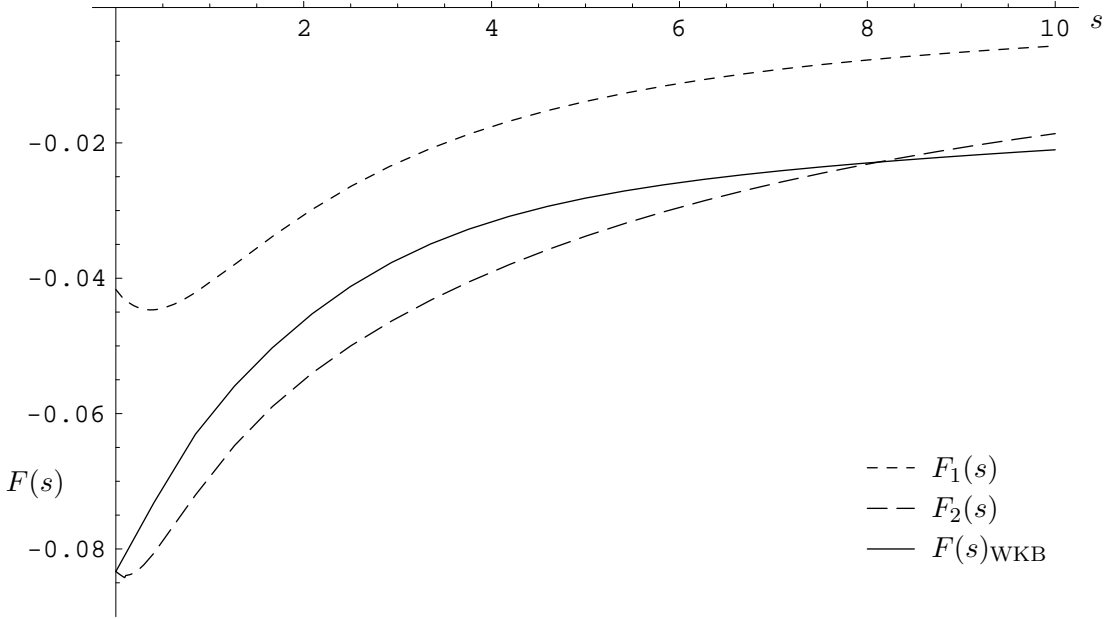}
\begin{picture}(40,20)(60,-30)
\put(0,30){$F_1 (s)$} \put(0,15){$F_2 (s)$} \put(0,0){$F(s)_{\rm
WKB}$} \put(60,200){$s$} \put(-350,25){$F(s)$}
\end{picture}
\caption{Plot of $F(s)$, showing the 1st, 2nd and 3rd order WKB approximations.}
\label{fig1}
\end{figure}

\begin{figure}[t]
\includegraphics{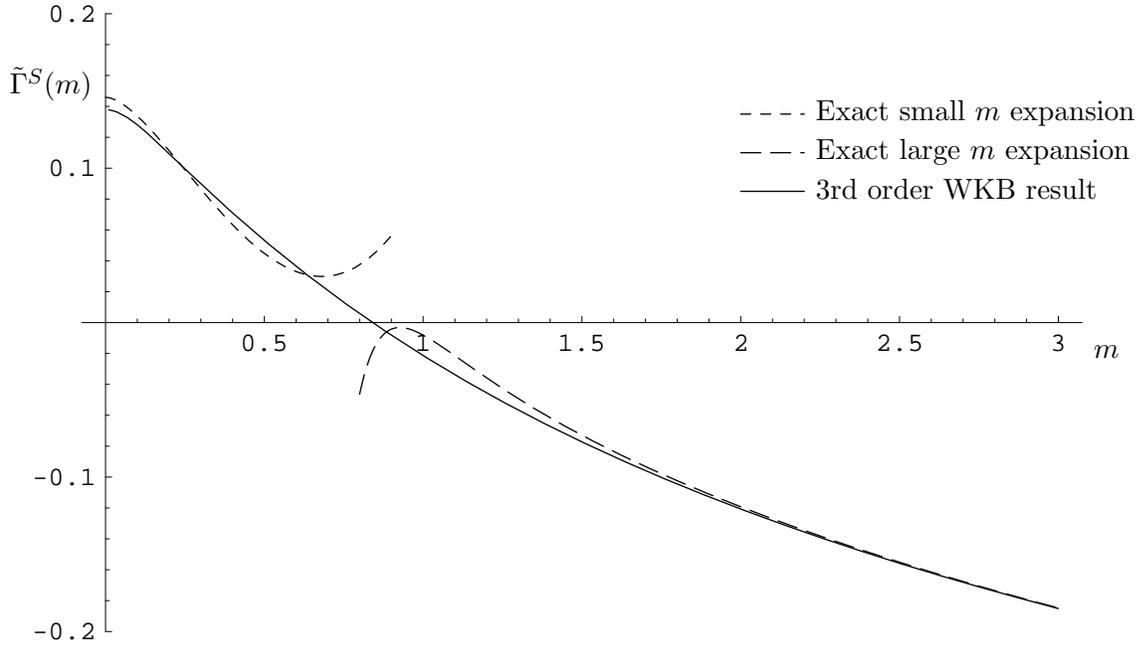}
\begin{picture}(40,20)(60,-30)
\put(-45,170){Exact small $m$ expansion} \put(-45,155){Exact large
$m$ expansion} \put(-45,140){3rd order WKB result} \put(60,80){$m$}
\put(-350,180){$\tilde{\Gamma}^{S}(m)$}
\end{picture}
\caption{Plot of $\tilde{\Gamma}^{S}(m)$, comparing the 3rd order WKB result with the exact extreme large and small mass limits.
\label{fig2}}
\end{figure}

Based on the form (\ref{local}), one can determine the function
$F(s)_{\rm WKB}$ numerically. The infinite $l$-sum in this formula
can be handled using the Euler-Maclaurin summation formula
\cite{abramowitz} (some care is needed 
for small-$r$ values), and
the result is a rapidly convergent series. Numerical integration
with the resulting function of $r$ can then be performed with very
high accuracy. Figure \ref{fig1} shows a comparison of the WKB
approximations for $F(s)$ with one, two and three terms in the WKB
expansion. We plot the function $F(s)_{\rm WKB}$ (i.e., including
up to 3rd-order WKB terms), together with the corresponding plots
for $F_1 (s) \equiv F^{(1)}(s)$ (i.e., the result based on the
leading WKB phase-shift expressions only) and $F_2 (s) \equiv
F^{(1)}(s)+F^{(2)}(s)$ (i.e., including up to 2nd-order WKB
corrections).
The value we found for $F_1 (s)$ at $s=0$ is $-\frac{1}{24}$, while the correct value is $- \frac{1}{12}$ (see (\ref{smalls})). Both 2nd and 3rd order WKB give the correct value for $F(0)$.
Thus, as remarked earlier, the leading WKB result
alone is not sufficient here even for the renormalization
discussion. With the 2nd order WKB expression, $F_2 (s)$, there is no problem with
renormalization, but, for its first derivative at $s=0$, we found
$F_2 '(0)=-  \frac{1}{90}$, which does not agree with the small $s$ result from the Schwinger-DeWitt expansion in  (\ref{smalls}). On the other hand, going to 3rd order WKB, $F(s)_{\rm
WKB}$ gives the correct value for this first
derivative, with $F'(0)_{\rm WKB}=\frac{1}{75}$. With the small-$s$ behavior satisfactorily taken
care of, we expect good agreement with the large mass behavior of the effective action, as can be confirmed from Figure \ref{fig2}.
For very large $s$ (which corresponds to the $m\to 0$ limit), the 3rd order WKB expression $F(s)_{\rm WKB}$ approaches zero faster than the true $F(s)$. Nevertheless, we show below that this 3rd order expression gives an excellent approximation even in the extreme massless limit.

Our function $F(s)_{\rm WKB}$ can be used to determine the
one-loop effective action for arbitrary mass value. We insert $F(s)_{\rm WKB}$
into (\ref{ptaction}), integrate over the proper-time $s$ (numerically), and
renormalize according to (\ref{renaction}). Extracting
$\tilde{\Gamma}^{S}(m)$, as defined in (\ref{modaction}), we obtain the plot shown in Fig. \ref{fig2}.
Also shown in Fig. \ref{fig2} are the curves based on the inverse
mass expansion (\ref{largemass}) and small mass expansion (\ref{smallmass}).
Evidently, our WKB-based plot corresponds to a smooth
interpolation of the latter two curves, as conjectured in
Ref.\cite{kwon} earlier. The agreement at large $m$ is excellent.
For $m=0$, our WKB-based prediction gives
rise to the value 0.137827, which is about 6\% off from 't Hooft's exact
value, $\alpha(\frac{1}{2})\simeq 0.145873$. [If $F_2 (s)$ were used
instead of $F(s)_{\rm WKB}$ for this calculation, the predicted
value for $\tilde{\Gamma}^{S}(0)$ would be 0.158084, which is
about 10\% off from the exact value]. The discrepancy from the
exact result is expected to be largest for $m=0$; hence, we estimate that our
WKB-based prediction of $\tilde{\Gamma}^{S}(m)$, for arbitrary
mass $m$, is good to 6\% accuracy.

\section{Field-theoretic derivative expansion approach}

The field-theoretic derivative expansion provides a quick and extremely simple estimate of the one-loop effective action in an instanton background for any value of the quark mass $m$, and we show here that even the leading order term gives surprisingly good agreement. The philosophy of the derivative expansion is to compute the  one-loop effective Lagrangian for a covariantly constant background field, which can be done exactly, and then perturb around this constant background solution. The leading order derivative expansion approximation for the effective action is obtained by first taking the (exact) expression for the effective Lagrangian in a covariantly constant background,  substituting the space-time dependent background, and then integrating over space-time.

For an instanton background, which is self-dual, we should base
our derivative expansion approximation on a covariantly constant
and self-dual background: one may then set (in a suitable
gauge)\cite{leutwyler} 
$ F_{\mu\nu}= F_{\mu\nu}^{\rm AB} n^a \, T^a\ , $ where the
abelian field strength $F_{\mu\nu}^{\rm AB}$ is self-dual, and
$n^a$ is a unit vector in color-space. Comparing with the large
$\rho$ limit of the $su(2)$ instanton field strength
(\ref{instf}), we identify $F_{\mu\nu}^{\rm
AB}n^a=-\frac{4}{\rho^2}\eta_{\mu\nu a}$. The exact one-loop scalar 
effective Lagrangian for a covariantly constant self-dual field is
well-known (see, e.g., Eq.
(2.11) in \cite{ds}). Substituting the
instanton form we obtain the derivative expansion (DE)
approximation (here we set $\rho=1$, as before)
\begin{eqnarray}
{\mathcal L}^{\rm scalar}_{\rm DE}=-\frac{2}{(4\pi)^2}\int_0^\infty \frac{ds}{s^3}\, e^{-m^2 s}\left[ \left(\frac{\frac{\sqrt{12} \, s }{(1+r^2)^2}}{\sinh\left(\frac{\sqrt{12}\, s}{(1+r^2)^2}\right)}\right)^2-1+\frac{1}{3}\left(\frac{\sqrt{12}\, s }{(1+r^2)^2}\right)^2   \right]
\label{sdlag}
\end{eqnarray}
The leading derivative expansion for the effective action is then obtained by integrating the effective Lagrangian 
(\ref{sdlag}) over space-time. Note that (\ref{sdlag}) has been renormalized on-shell, so that $\mu=m$ in (\ref{modaction}). 
To study both the large and small mass limits it is useful to express (\ref{sdlag}) in a different, but equivalent, form using the identity:
\begin{eqnarray}
\int_0^\infty \frac{du}{u^3}e^{-2 \kappa u}\left[\left(\frac{u}{\sinh u}\right)^2-1+\frac{u^2}{3}\right]= -\frac{1}{3}\ln \kappa +4\int_0^\infty \frac{dx\, x}{e^{2\pi x}-1}\, \ln\left(x^2+\kappa^2\right)
\label{id}
\end{eqnarray}
Then the large mass expansion of the effective action is straightforward:
\begin{eqnarray}
\Gamma^S_{\rm DE}(A; m)&=&2\pi^2\int_0^\infty r^3\, dr\, {\mathcal L}^{\rm scalar}_{\rm DE}\nn\\
&=&-6\int_0^\infty \frac{r^2\, d(r^2)}{(1+r^2)^4}\int_0^\infty \frac{dx\, x}{e^{2\pi x}-1}\, \ln \left(1+\frac{48 x^2}{m^4(1+r^2)^4}\right)\nn\\
&\sim& 3\sum_{l=1}^\infty \frac{48^l 
\, {\mathcal
B}_{2l+2}}{(2l+2)(2l+1)(2l)(4l+3)} \frac{1}{m^{4l}}\quad, \quad  m
\to \infty \label{dlm}
\end{eqnarray}
where ${\mathcal B}_{l}$ are the Bernoulli numbers. Note that the large mass expansion (\ref{dlm}) of the leading derivative expansion approximation begins with $\frac{1}{m^4}$, rather than the true $\frac{1}{m^2}$ behavior, because for the covariantly constant self-dual field the corresponding Schwinger-DeWitt coefficient vanishes by group theoretic traces.

\begin{figure}[t]
\includegraphics{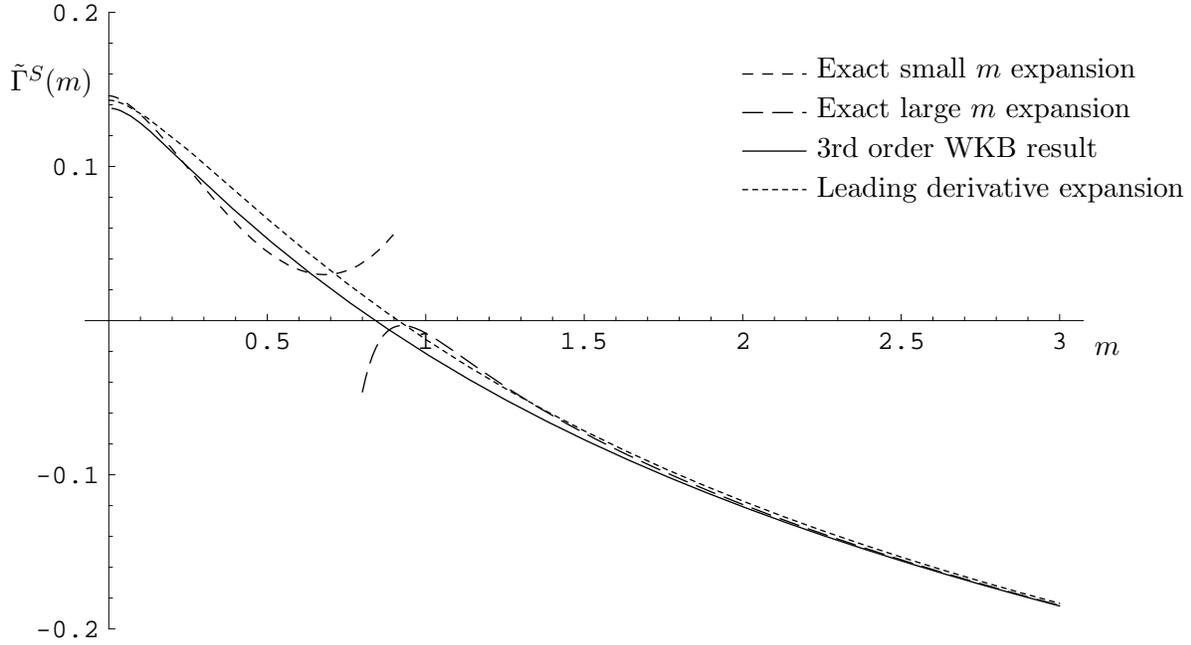}
\begin{picture}(40,20)(60,-30)
\put(-45,185){Exact small $m$ expansion} \put(-45,170){Exact large
$m$ expansion} \put(-45,155){3rd order WKB result} \put(-45,140){Leading derivative
expansion} \put(60,80){$m$}
\put(-350,180){$\tilde{\Gamma}^{S}(m)$}
\end{picture}
\caption{Plot of $\tilde{\Gamma}^{S}(m)$, comparing the leading derivative expansion approximation with the 3rd order WKB result and
with the exact extreme large and small mass limits. \label{fig3}}
\end{figure}

To study the small mass expansion, we perform the $r$ integral in (\ref{dlm}) to obtain an exact integral representation of the leading derivative expansion approximation
\begin{eqnarray}
\Gamma^S_{\rm DE}(A; m)
&=&-\frac{1}{14}\int_0^\infty \frac{dx\, x}{e^{2\pi x}-1}\left\{ -84+14 \ln\left(1+\frac{48 x^2}{m^4}\right)
+7\sqrt{3}\,\frac{m^2}{x} {\rm arctan}\left(\frac{4\sqrt{3}\, x}{m^2}\right) \right.\nn\\
&&\left. \hskip 3cm +768\frac{x^2}{m^4}~_2 F_1\left(1,\frac{7}{4},\frac{11}{4};-\frac{48x^2}{m^4}\right)\right\}
\label{dm}
\end{eqnarray}
It is now simple to expand each term for small mass to obtain the leading behavior
\begin{eqnarray}
\Gamma^S_{\rm DE}(A; m)\sim  \frac{1}{6}\ln (m )+\left(\frac{5}{36}-\frac{1}{24}\ln(48)-\zeta^\prime(-1)\right)+\frac{\sqrt{3}}{4}\,m^2\, \ln (m)+\dots
\,\,, \,\, m \to 0
\label{dsm}
\end{eqnarray}
The agreement with the leading small mass behavior
(\ref{smallmass}) of the exact result is quite remarkable. The
coefficient $\frac{1}{6}$ of the $\ln(m)$ term agrees, as it must
by virtue of the $\beta$-function. The constant term
$\left(\frac{5}{36}-\frac{1}{24}\ln(48)-\zeta^\prime(-1)\right)\simeq
0.14301$ is only $2\%$ away from 't Hooft's value of
$\alpha(\frac{1}{2})\simeq 0.14587$, and the coefficient of the
$m^2 \ln (m)$ term is $\frac{\sqrt{3}}{4}\simeq 0.433$, compared
to Carlitz and Creamer's result of $0.5$. Figure \ref{fig3}
 shows a comparison of the associated function $\tilde{\Gamma}^S(m)$ with the WKB result of the previous section and the exact large and small mass limits. The agreement is surprisingly good for such a crude approximation \cite{diakonov}.

\section{Conclusion}

In this letter we have presented a computation of the fermion determinant in an instanton background for all values of the quark mass. Using 3rd order WKB approximation for the scattering phase shifts we obtained a result which interpolates very well between the known extreme large and small mass results. We expect even higher order WKB terms to improve the accuracy further. Also, the agreement in the small mass regime might be improved by determining the large $s$ asymptotic behavior of the proper-time function $F(s)$ generalizing the method of Barzinsky and Mukhanov \cite{barvinsky}. Finally, we showed that a very crude leading order derivative expansion approximation based on a covariantly constant self-dual background leads to surprisingly good agreement.

\vskip .5cm
{\bf Acknowledgments:} GD thanks D. Diakonov and H. Gies for helpful comments,
the US DOE for support through the grant DE-FG02-92ER40716, and the CSSM at Adelaide for hospitality.
The work of CL was supported by the Korea Science Foundation ABRL
program (R14-2003-012-01002-0) and by the Korea Research
Foundation Grant 2001-015-DP0085.


\begin{thebibliography}{12345}

\bibitem{heisenberg} W. Heisenberg and H. Euler, Z. Phys.
\textbf{98}, 714 (1936); V. Weisskopf, Kong. Dans. Vid. Selsk.
Math-fys. Medd. XIV No. 6 (1936).

\bibitem{schwinger} J. Schwinger, Phys. Rev. \textbf{82},
664 (1951).

\bibitem{brownduff}
M.~R.~Brown and M.~J.~Duff,
Phys.\ Rev.\ D {\bf 11}, 2124 (1975).

\bibitem{dunne} For a recent review, see: G.~V.~Dunne,
``Heisenberg-Euler effective Lagrangians: Basics and extensions,''
arXiv:hep-th/0406216,
To appear in Ian Kogan Memorial Collection, {\it From Fields to Strings: Circumnavigating Theoretical Physics}, M. Shifman et al (Eds), (World Scientific, Singapore).

\bibitem{thooft} G. 't Hooft, Phys. Rev. \textbf{D14}, 3432 (1976); {\it ibid}
\textbf{18}, 2199 (E) (1978).

\bibitem{brown} L. S. Brown and D. B. Creamer, Phys. Rev.
\textbf{D18}, 3695 (1978); E. Corrigan, P. Goddard, H. Osborn, and
S. Templeton, Nucl. Phys. \textbf{B159}, 469 (1979); B. Berg and M.
L\"{u}scher, Nucl. Phys. \textbf{B160}, 281 (1979); H. Osborn, Ann.
Phys.(N. Y.) \textbf{135}, 373 (1981);
C.~Lee, H.~W.~Lee and P.~Y.~Pac,
Nucl.\ Phys.\ B {\bf 201}, 429 (1982).

\bibitem{belavin} A. Belavin, A. Polyakov, A. Schwartz, and Y.
Tyupkin, Phys. Lett. \textbf{59B}, 85 (1975).

\bibitem{schaffer} T. Sch\"{a}fer and E. V. Shuryak, Rev. Mod.
Phys. \textbf{70}, 323 (1988); M.~A.~Shifman,
World Sci.\ Lect.\ Notes Phys.\  {\bf 62}, 1 (1999);
D. Diakonov, Prog. Part. Nucl. Phys. {\bf 5}, 173 (2003).


\bibitem{lattice}
A.~W.~Thomas,
Nucl.\ Phys.\ Proc.\ Suppl.\  {\bf 119}, 50 (2003)
[arXiv:hep-lat/0208023];
C.~Bernard et al,
Nucl.\ Phys.\ Proc.\ Suppl.\  {\bf 119}, 170 (2003)
[arXiv:hep-lat/0209086].

\bibitem{kwon} O-K. Kwon, C. Lee and H. Min, Phys. Rev.
\textbf{D62}, 114022 (2000).

\bibitem{dewitt} B. S. DeWitt, "Dynamical Theory of Groups and
Fields" (Gordon and Breach, New York, 1965);  Phys. Rep.
\textbf{19}, 295 (1975).

\bibitem{lee} C. Lee and C. Rim, Nucl. Phys. \textbf{B255},
439(1985); I. Jack and H. Osborn, ibid. \textbf{B249}, 472 (1985);
R. D. Ball, Phys. Rep. \textbf{182}, 1 (1989).

\bibitem{shifman}
V.~A.~Novikov, M.~A.~Shifman, A.~I.~Vainshtein and V.~I.~Zakharov,
Fortsch.\ Phys.\  {\bf 32}, 585 (1985).


\bibitem{carlitz} R. D. Carlitz and D. B. Creamer, Ann. Phys.(N.
Y.) \textbf{118}, 429 (1979).

\bibitem{wasson} D. A. Wasson and S. E. Koonin, Phys. Rev.
\textbf{D43}, 3400 (1991); C. L. Y. Lee, Phys. Rev. \textbf{D49},
4101 (1994).

\bibitem{moss} I. G. Moss, Phys. Lett. \textbf{B460}, 103 (1999);
I. G. Moss and W. Naylor, Nucl. Phys. \textbf{B632}, 173 (2002).

\bibitem{dunham} J. L. Dunham, Phys. Rev. \textbf{41}, 713 (1932);
C. Rosenzweig and J. B. Krieger, J. Math. Phys. \textbf{9},
849 (1968); C. M. Bender, Phys. Rev. \textbf{D16}, 1740 (1977).

\bibitem{bender} C. M. Bender and S. A. Orszag, {\it Advanced
Mathematical Methods for Scientists and Engineers}, (McGraw-Hill
Inc., New York, 1978), Sec. 10. 7.

\bibitem{langer} R. E. Langer, Phys. Rev. \textbf{51}, 669 (1937);
J. B. Krieger and C. Rosenzweig, Phys. Rev. \textbf{164},
171 (1967).

\bibitem{preparation} G. V. Dunne, Jin Hur, C. Lee, and H. Min, in
preparation.

\bibitem{abramowitz} M. Abramowitz and I. Stegun, "Handbook of
Mathematical Functions"(Dover, New York, 1965).

\bibitem{leutwyler} H. Leutwyler, Phys. Lett. \textbf{B96},
154 (1980); Nucl. Phys. \textbf{B179}, 129 (1981).

\bibitem{ds} G.~V.~Dunne and C.~Schubert,
JHEP {\bf 0208}, 053 (2002)
[arXiv:hep-th/0205004].

 \bibitem{diakonov} Similar surprisingly good agreement of the derivative expansion approximation
for self-dual backgrounds has been found by D. Diakonov: private communication.

\bibitem{barvinsky}

A.~O.~Barvinsky and V.~F.~Mukhanov,
Phys.\ Rev.\ D {\bf 66}, 065007 (2002)
[arXiv:hep-th/0203132].



\end{thebibliography}
\end{document}